\documentstyle[cite,graphicx]{article}

\def \eqtext#1		{\hspace{1in} \hbox{#1}}	
\def \micron    	{$\mu$m}

\def \idlplot#1		{\centerline{\scalebox{0.9}{\includegraphics{#1}}}} 
\def \idlplotps#1	{\centerline{\scalebox{0.9}{\includegraphics[70,350][574,710]{#1}}}} 

\def \subsimt#1{{\lower 2pt\hbox{$\scriptstyle #1$}\atop
     \raise 1pt\hbox{$\scriptstyle \sim$}}}

 1
 2

 1
 2

\def \Ha {H$\alpha$}
\def \Paa {Paschen~$\alpha$}
\def \arcsec {"}

\begin{document}
\centerline {\bfseries \Large Evidence for Dust Grain Growth in Young Circumstellar Disks}
\centerline{
  Henry B. Throop\footnote{Campus Box 392 / LASP, University of Colorado,
  Boulder, CO 80309-0392. Present address: Southwest Research Institute,
  1050 Walnut St. Ste. 426, Boulder, CO 80302.  throop@boulder.swri.edu},
  John Bally, 
  Larry W. Esposito, 
  Mark J. McCaughrean}
\centerline{Submitted to {\sl Science}: 16 January 2001}
\centerline{Revised: 19 April 2001}

\ \\

{\bf 
Hundreds of circumstellar disks in the Orion nebula are being rapidly
destroyed by the intense ultraviolet radiation produced by nearby bright
stars.  These young, million-year-old disks may not survive long
enough to form planetary systems.  Nevertheless, the first stage of
planet formation -- the growth of dust grains into larger particles
-- may have begun in these systems.  Observational evidence for
these large particles in Orion's disks is presented.  A model of grain
evolution in externally irradiated protoplanetary disks is developed and
predicts rapid particle size evolution and sharp outer disk boundaries.
We discuss implications for the formation rates of planetary systems.}

The growth of dust grains orbiting young stars represents the first
stage of planet formation.\cite{bhn00}  However, stars born in massive
star-forming regions such as the Orion nebula are heated by intense
ultraviolet (UV) radiation from nearby O and B stars, and the gas and
dust in their disks can be lost in less than $10^5\rm\ years$.\cite{ho99}
Planet formation in such environments may therefore be inhibited if
it requires substantially longer than this time.\cite{sh98b} But,
if growth to large particles can occur before removal of the gas and
small particles, planets may nevertheless form from these disks.  In this
paper, visual and near-infrared wavelength images obtained with the
Hubble Space Telescope (HST) are used to show that particles in Orion's
largest disk have grown to radii larger than 5~\micron .  Furthermore,
the absence of millimeter-wavelength emission may provide evidence that
grains have grown to sizes larger than a few millimeters.  We develop
a grain evolution model incorporating the effects of photo-ablation
that demonstrate that the timescale for grain growth can be shorter
than the photo-evaporation time.  It is thought that the majority of
stars in the Galaxy form in photo-evaporating regions like the Orion
nebula;\cite{wan00} if this is true, then giant planets and Kuiper belts
of icy bodies around stars are probably rare unless they are formed
very rapidly.

Solar system-sized circumstellar disks in the Orion nebula were first
inferred from radio observations of dense ionized regions surrounding
young low-mass stars.\cite{cfw87}  HST subsequently yielded images of
extended circumstellar material surrounding over half of the observed
300 young low-mass stars in core of the Orion nebula\cite{owh93,mo96}.
Most of these `proplyds' consist of comet-shaped ionized envelopes
pointing directly away from the brightest stars in the Nebula
\cite{mcb98,bom00}.   Proplyds are believed to contain evaporating
circumstellar disks\cite{jhb98} and over 40 disks have been resolved on
HST images.  More than 25 are found inside ionized envelopes while 15
are seen purely in silhouette against the background light of the nebula.

Assuming disk masses $\sim 0.01-0.05 M_\odot$ ($1 M_\odot = 1$ solar
mass)\cite{jhb98}, external radiation erodes disks in the central 1~pc of
the Orion nebula on $10^4$ to $10^5 \rm\ yr$ timescales\cite{jhb98,sh98b}.
Soft UV photons (91~nm $< \lambda <$ 200~nm) from nearby massive stars
heat the disk surface layers to about 1000~K.  Gas heated above the
local escape velocity is lost from the disk at loss rates of $\dot M
\approx 10^{-7}$ to $10^{-6} M_\odot\rm\ yr^{-1}$ \cite{jhb98,ho99} and
forms the cometary `proplyds' surrounding many of Orion's young stars.
Dust grains will be entrained in the escaping neutral outflow where the
gas drag forces on them exceed the force of gravity; the small ionized
outflow component has negligible effect on grain loss.\cite{thr00}
Entrained dust has been observed just inside the ionization fronts in
several proplyds,\cite{bom00} but not considered in previous modeling.


The properties of the grains in Orion's circumstellar disks can
be probed  by the wavelength-dependence of the attenuation of the
background nebular light that filters through the translucent disk edges.
Standard interstellar\footnote{
  The term `interstellar' refers to the population of small, primordial
  dust grains in the Orion nebula that have not been processed in a
  circumstellar disk.  }
grains with radii of 0.1-0.2~\micron\cite{kmh94} scatter shorter
wavelength visible light more efficiently than longer wavelengths.
Therefore, disks containing predominantly small interstellar grains
become more transparent with increasing wavelength.  On the other hand,
the opacity of disks containing predominantly large grains (larger than
several times the wavelength) will be independent of the wavelength.
While small grains `redden' transmitted light, large grains do not alter
its color, rendering the translucent portion of the disk `grey.'

The largest circumstellar disk in the Orion nebula,  114-426, is seen in
silhouette against background nebular light.  This nearly edge-on disk
(the central star is occulted by the disk) has a radius of $\sim$ 550~
astronomical units (AU) and a resolved translucent outer edge roughly
200 by 200~AU in size at its northeast ansa (Fig.~1).\cite{bom00}
Grain properties in this region can be probed by the attenuation of the
bright background 1870~nm \Paa\ and 656~nm \Ha\ lines.  Because both
these lines originate from recombinations of ionized hydrogen, the
brightness ratio between these two lines is relatively constant over
the extent of 114-426.

We used the Planetary Camera of HST's WFPC2 instrument to obtain a
set of four dithered 400 second \Ha\ exposures of 114-426 on January
11, 1999, resulting in reduced images with an angular resolution
of 0.07\arcsec, or 30~AU.\cite{bom00}  We also used HST's NICMOS1
camera to obtain a 640 second \Paa\ exposure on 26 February 1998 at
resolution of 0.16\arcsec.\cite{msc00}  In order to compare these images
at identical resolutions, we convolved the \Ha\ image with a synthetic
\Paa\ point-spread function (PSF), and the \Paa\ image with an \Ha\ PSF.

Linear slices through the disk midplane at 656~nm and 1870~nm show that
the opacity profiles of the translucent western edge of 114-426 are
indistinguishable (Fig.~2).  Thus, background light is not `reddened';
dust at the disk edge is `grey' to a level of $\sim 5\%$ between 656 nm
and 1870~nm.  The mean extinction of the translucent ansa of the 114-426
disk at these wavelengths can be compared to the reddening produced by
observations of interstellar grains along several typical lines-of-sight
(Fig. 3).  The 114-426 disk's translucent edge is achromatic and can
not be fit by any standard interstellar extinction law.  The standard
interstellar extinction curve indicates that the opacity should be
5-10 times lower at 1870~nm than at 656~nm.  This result depends only
weakly on composition, grain shape, or the presence of fractal aggregates
\cite{lrh97,mtm96}.  The observed grey opacity indicates that extinction
is dominated by particles larger than 5~\micron\ in radius, 25-50 times
larger than typical interstellar grains.  In contrast to the NE ansa, the
disk's polar halo region decreases in size with wavelength, indicating a
suspended population of small particles above the disk poles.\cite{msc00}
Although previous observations of 114-426\cite{mcb98} indicated a
decrease in disk size by $20\%$ from 656~nm to 1870~nm, that result
may have reflected poor signal/noise ratio in the earlier 1870~nm
observations\cite{thr00}.

The lack of millimeter-wavelength emission places additional constraints
on grain sizes.  We observed six Orion nebula disks, including 114-426,
with the Owens Valley Radio Observatory (OVRO) millimeter wavelength
interferometer at $\lambda$ = 1.3~mm continuum.\cite{bts98}  None were
detected, implying mass limits of $M_{\rm disk} < 0.020 M_\odot$ under the
assumption of an interstellar emissivity of $2\times 10^{-2}\ \rm{cm^{2}\
g^{-1}}$\cite{mll95}.  However, in re-visiting our analysis of these
data, we note the possibility that grains have grown larger than a few
millimeters and thus the standard emissivity may underestimate the mass.
The models described below predict that particles grow to larger than 1
mm throughout the disk in less than $10^5\rm\ yrs$, causing an emissivity
of $2\times 10^{-3}\rm\ cm^2\ g^{-1}$ and implying disk masses as large
as 0.2 $M_\odot$.

%
%
%
%
%
%
%
%
%

We have developed a numerical model\cite{thr00} to explore grain behavior
within photo-evaporating disks.  Our model includes (i) grain growth due
to mutual collisions and accumulation of ice mantles,\cite{mmv88} (ii)
coupling and loss of small grains entrained in UV photo-evaporation
induced outflow,\cite{jhb98,sh98b} and (iii) photosputtering of
ices.\cite{wbj95}  The grain density is assumed to remain constant
as grains collide and stick within turbulent eddies produced by heat
escaping from the disk midplane.  Vertical and azimuthal symmetry is
assumed\cite{dms95} and the abundance and size distribution of ice,
silicate, and gas are tracked separately at each time-step.  Our disk
has an initial mass of 0.1 $M_{\odot}$, with a grain size
distribution identical to interstellar dust.  The model starts when the
ionizing source turns on, and stops when the disk thermal optical depth
has dropped below unity and can no longer sustain convection,\cite{mmv88}
typically in a little over $10^5 \rm\ yr$.  After convection stops,
grain growth is dominated by processes such as settling and gravitational
interactions. These processes are not considered here.  The model
does not simulate 114-426 in its current, non-photo-evaporating
state\footnote{114-426 is thought to show no photo-evaporation because
it is
  outside the Orion core's Str\"omgren sphere and thus receives no soft UV
  flux; we note the possibility that it may not be photo-evaporating today
  because all gas has already been lost in previous photo-evaporative
  episodes, and it is a pure dust disk.};
rather, it simulates the 114-426 disk as it would appear placed near
the Orion nebula core at a distance 0.1~pc.


Grains grow most rapidly in the center of the model disk (Fig. 4a) where
the highest densities and temperatures are found and photo-evaporation
does not operate.  The growth rate decreases with distance from the
central star;  grain sizes reach $r=1\rm\ m$ at 10~AU and $r=1\rm\
mm$ at 500~AU in $10^5\rm\ yr$.  In the outer disk, small ($<$ 1
mm) grains are entrained in the photo-evaporative flow from the disk
surface and lost from the system, decreasing the optical depth (Fig. 4b).
After $10^5\rm\ yr$, few particles remain in the disk outward of
40~AU.  At the transition between these `grain growth' and `grain loss'
regimes, an edge populated by large, cm-sized particles is left behind.
After the silicate population has stabilized, photo-sputtering continues
to reduce ice particle sizes and remove gas, and nearly all ice and gas
are removed by $10^6\rm\ yr$.  Only silicates that grow large enough
($r~>$~1~mm) to resist photo-evaporative entrainment are retained.
Ices do not survive and only rocky planets, planetary cores, or asteroids
can form.

In the standard planetary formation model, giant planets such as
Jupiter form by accreting hydrogen and helium rich gas from the disk
onto a large rocky core.\cite{phb96}  In Orion-like environments,
there may not be time to grow the requisite cores before loss of
the gas since Jupiter's formation would require $10^6-10^7\rm\ yr$.
If giant planets form in Orion-like systems, they must do so before disk
photo-evaporation.  One viable mechanism for rapid formation of giant
planets is gravitational collapse, which has been postulated to occur in
disks with $M_{disk} > 0.13 M_{\odot}$ on $10^3$ year time-scales around
solar-mass stars.\cite{bos97}  Icy Kuiper belt objects and comets are
believed to have formation time-scales of $10^8-10^9\rm\ yr$.\cite{fds00}
Thus, these objects are also difficult to form in Orion-like environments.
The architectures of any new planetary systems that might form in Orion
are likely to be different from that of our Solar system.


The evidence for large particles in 114-426 complements several previous
studies of particles in young disks.  The reflected-light near-infrared
spectrum of the disk orbiting HR4796A\cite{ssb99} provides evidence
for particles with radii larger than 2-3~\micron .  Several disks
in NGC2024{\cite{vrc98} and Taurus\cite{me94} reveal relatively flat
sub-millimeter spectra that may indicate large grains.  However, near-IR
observations of the disk in HH30\cite{wsk01} show normal dust opacities
and no evidence for grain growth and sub-mm observations of the HL Tau
disk\cite{bhn00} are inconclusive.  Grain growth in disks appears to
depend strongly on their environment.

The majority of young stars in the Milky Way galaxy appear to have formed
in large, dense clusters like the Orion nebula, rather than in smaller,
dark clouds such as Taurus-Auriga.\cite{wan00}  Within large clusters,
the majority of stars form near massive stars where their disks can
be rapidly destroyed.   Thus, planet formation models must be revised
to consider the destructive effects of these environments.  We present
evidence for large gains in one Orion disk.  This creates the possibility
that planetary system with architectures different from our own Solar
system may nonetheless form in such hazardous environments.

\bibliography{1059093.bib}

{\bf Acknowledgments:} This research was funded by the Cassini
project, the NASA Astrobiology Institute, NASA grants NAG5-8108 and
GO-06603.01-95A, DLR grant number 50--OR--0004 and European Commission
Research Training Network RTN1--1999--00436.  We thank C. Campbell, N.
Turner, and two anonymous reviewers for their comments.

\pagebreak

{\bf Figure 1:} Images of the 114-426 disk in Orion at 656 nm (left)
and 1870 nm (right) obtained with HST.  The images have been rotated and
scaled to the same spatial scale, and are shown at full resolution before
convolving as described in the text.  The maximum and minimum background
intensities have been normalized to unity and zero respectively.
Both images were processed and calibrated through the standard HST
data pipeline.

{\bf Figure 2:} One-dimensional intensity slices along the major axis
of the 114-426 circumstellar disk.  Both images have been convolved
with their complementary PSF's to produce images at matched angular
resolutions.  The disk's SW ansa is consistent with a sharp disk edge.
In contrast, the disk's translucent NE ansa is spatially extended over at
least 4 resolution elements.  The indistinguishable profiles at the two
wavelengths indicates that transmission through the translucent portion
of the disk is achromatic, and the disk is dominated by particles larger
than 5\micron.

{\bf Figure 3:} The wavelength dependence of the light transmitted through
the translucent portion of the 114-426 disk, compared to the range of
standard observed interstellar reddening laws.  The 656 and 1870~nm
data points for the NE ansa of 114-426 is marked.  All data have been
normalized to the V photometric band.  The points for 114-426 show that
the extinction is grey and can not be represented by any interstellar
extinction law.   The error bar on the 1870~nm point shows 3~$\sigma$
errors, dominated by spatial variations in the background light at both
wavelengths, and flat-field artifacts in the 1870~nm image.  $A$ indicates
the extinction in magnitudes at wavelength $\lambda$.  The color ratio
$R_{\hbox{v}}$ is defined as $A_{\hbox{v}}/(A_{\hbox{v}} - A_{\hbox{B}})$.
Other single letters indicate the standard photometric bands.  The stellar
data were taken from \cite{ccm89}.

{\bf Figure 4:}  A model for grain growth in a photo-evaporating
circumstellar disk exposed to a UV radiation field typical of the Orion
Nebula.  The initial grain size distribution is that of interstellar
dust.  {\bf a)} The evolution of the particle size with time at
several radial distances. Solid lines correspond to radial distances
$R_i = \exp(2.31+0.43n_i)\rm\ AU$.  The outermost several bins do not
grow significantly and their lines appear superimposed on each other.
Particles grow quickly at the inner edge due to high collision velocities,
high gas densities, and slow loss processes.  Growth is terminated
when infrared optical depth drops below unity, inhibiting convection.
{\bf b)} The evolution of the radial profile of the disks opacity as
viewed from the disk axis.  Grains at the outer edge are removed by
the photo-ablation flow while grains in the inner disk grow rapidly.
These two processes create a disk populated by large particles.
Time-steps for the solid lines correspond to times $t_i = \exp(5.1 +
0.24n_i)\rm\ yr$.  The input parameters used here are representative
of photo-evaporative conditions 0.1~pc from the Orion nebula
core.\cite{jhb98,sh99}

The following disk parameters are assumed: The surface density
declines with disk radius as $\Sigma \sim  R^{-3.5}$\cite{ybl93}, the
vertical height scales as $z = R/10$, the sputtering rate is given by
$(dr/dt)_{\rm s} = 1\ \mu\rm m\ yr^{-1}$, and the sticking efficiency is
$\epsilon=0.1$. The outflow column density is $n_0 = 3\times 10^{22}\rm
cm^{-2}$, higher than the $3\times 10^{21}\rm\ cm^{-2}$ of \cite{sh99}
to account for deeper UV penetration due to the large grains in the disk.
The viscosity parameter is $\alpha = 10^{-2}$, the outflow temperature
at it base is $T_{\rm o} = 1000 K$, the central star mass is $M_{\rm s}
= 1 M_\odot$, the disk inner and outer radii are $R_{in}$ = 10 AU and
$R_{out}$ = 500 AU, and the grain density is $\rho=1\rm\ g\ cm^{-3}$.

\pagebreak

{\bf Figure 1:}
\ \ \\
\ \ \\

\centerline{\includegraphics[width=6in]{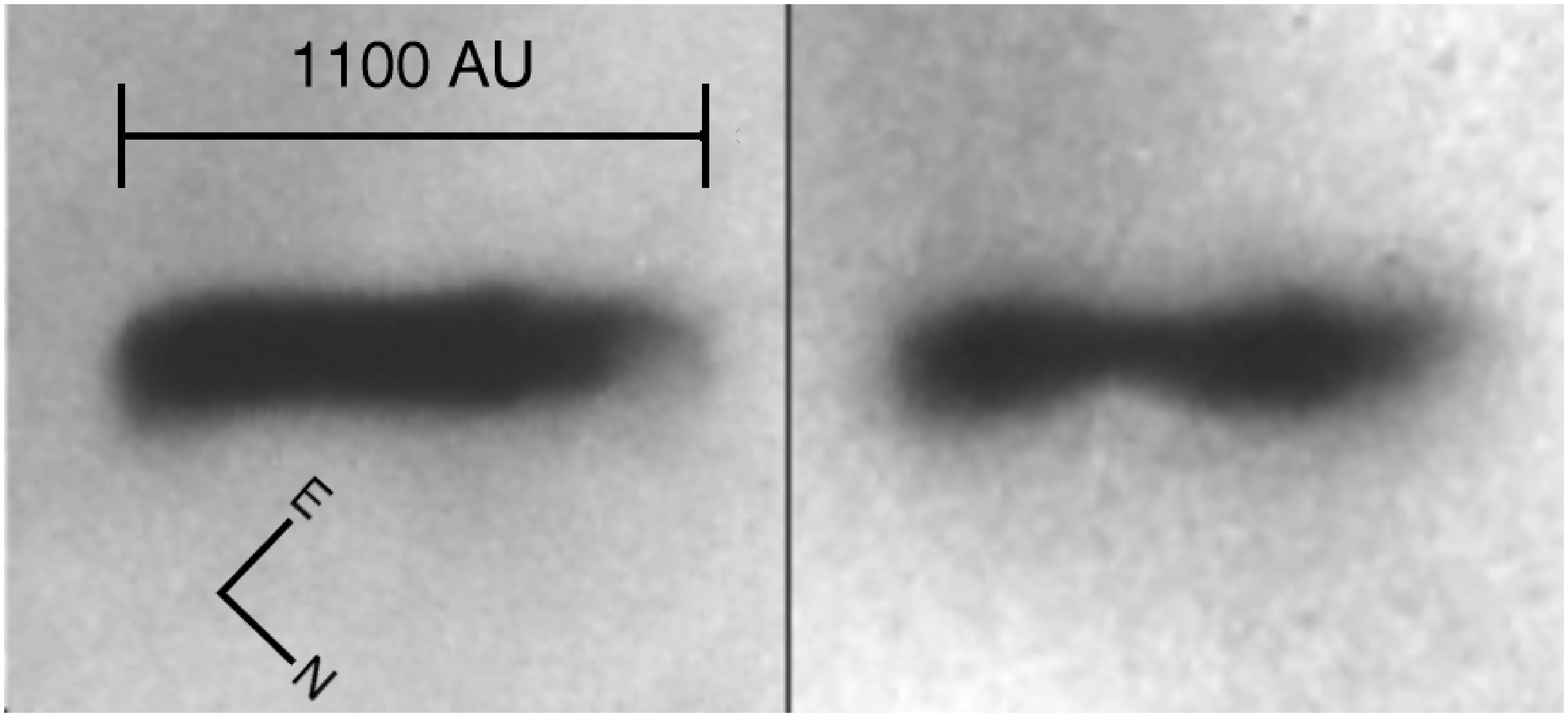}}
\pagebreak

{\bf Figure 2:}
\ \ \\
\ \ \\

\idlplot{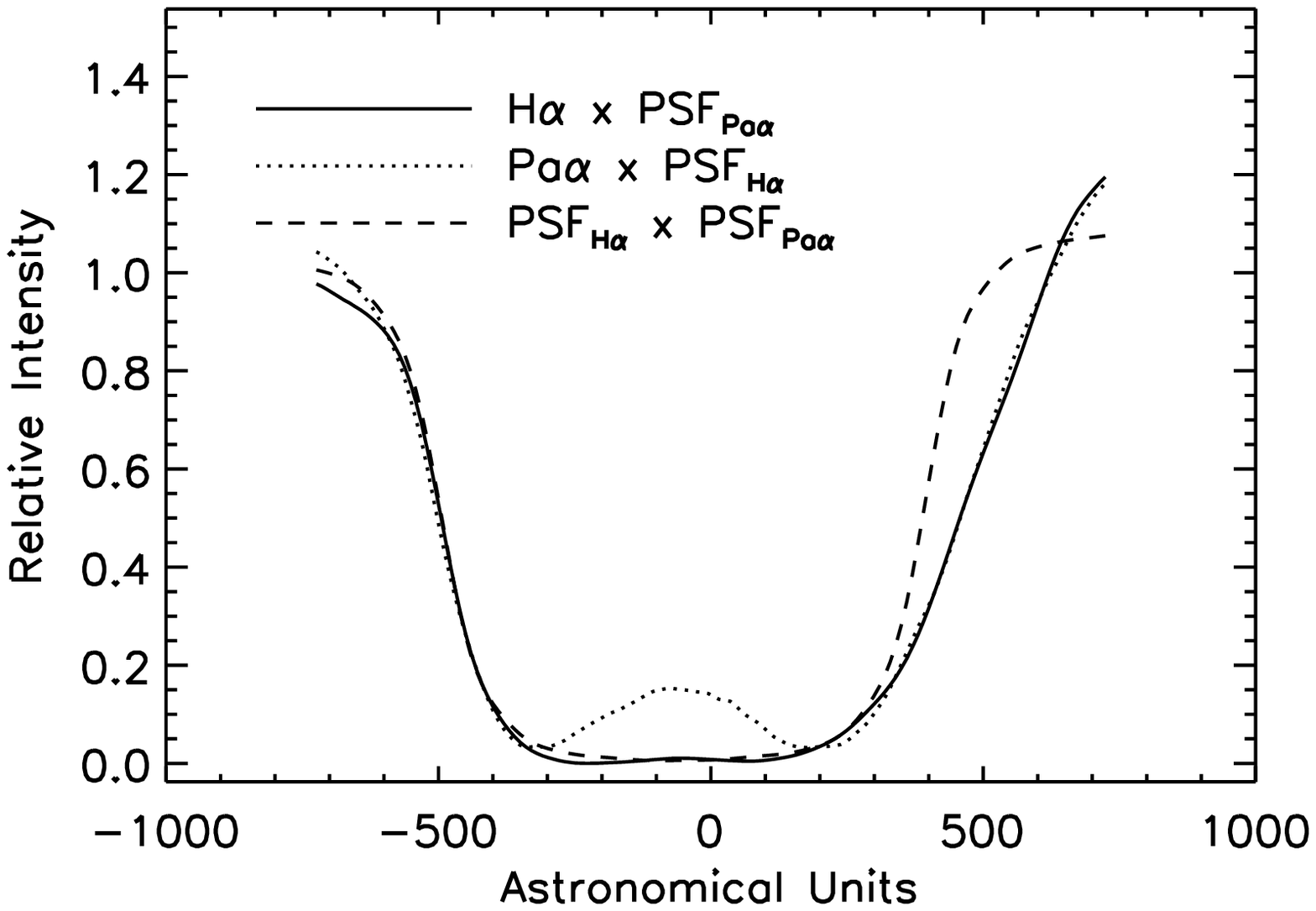}
\pagebreak

{\bf Figure 3:}
\ \ \\
\ \ \\

\scalebox{0.90}{\includegraphics{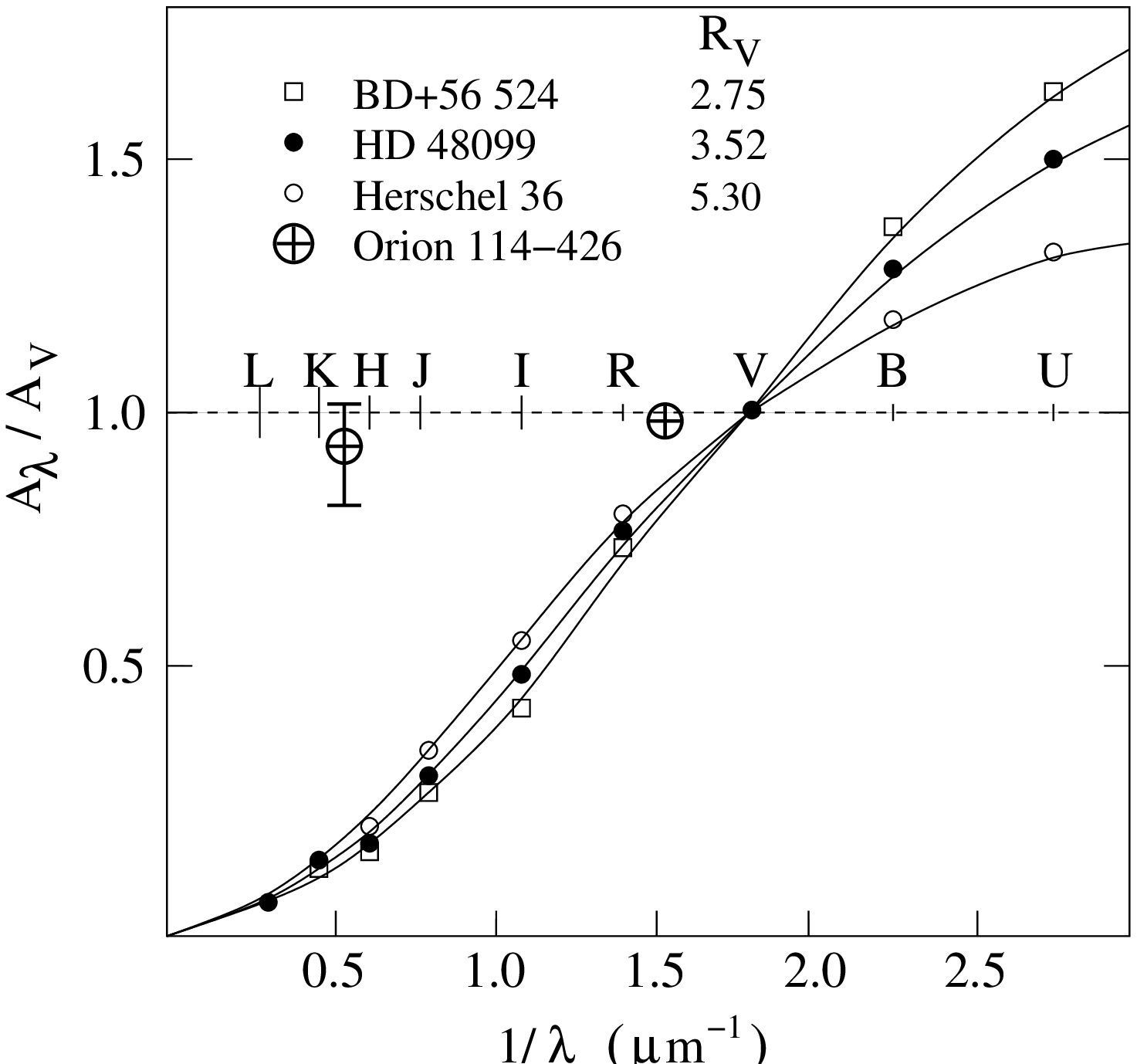}}

\pagebreak

{\bf Figure 4:}\\

\centerline{\scalebox{0.65}{\includegraphics{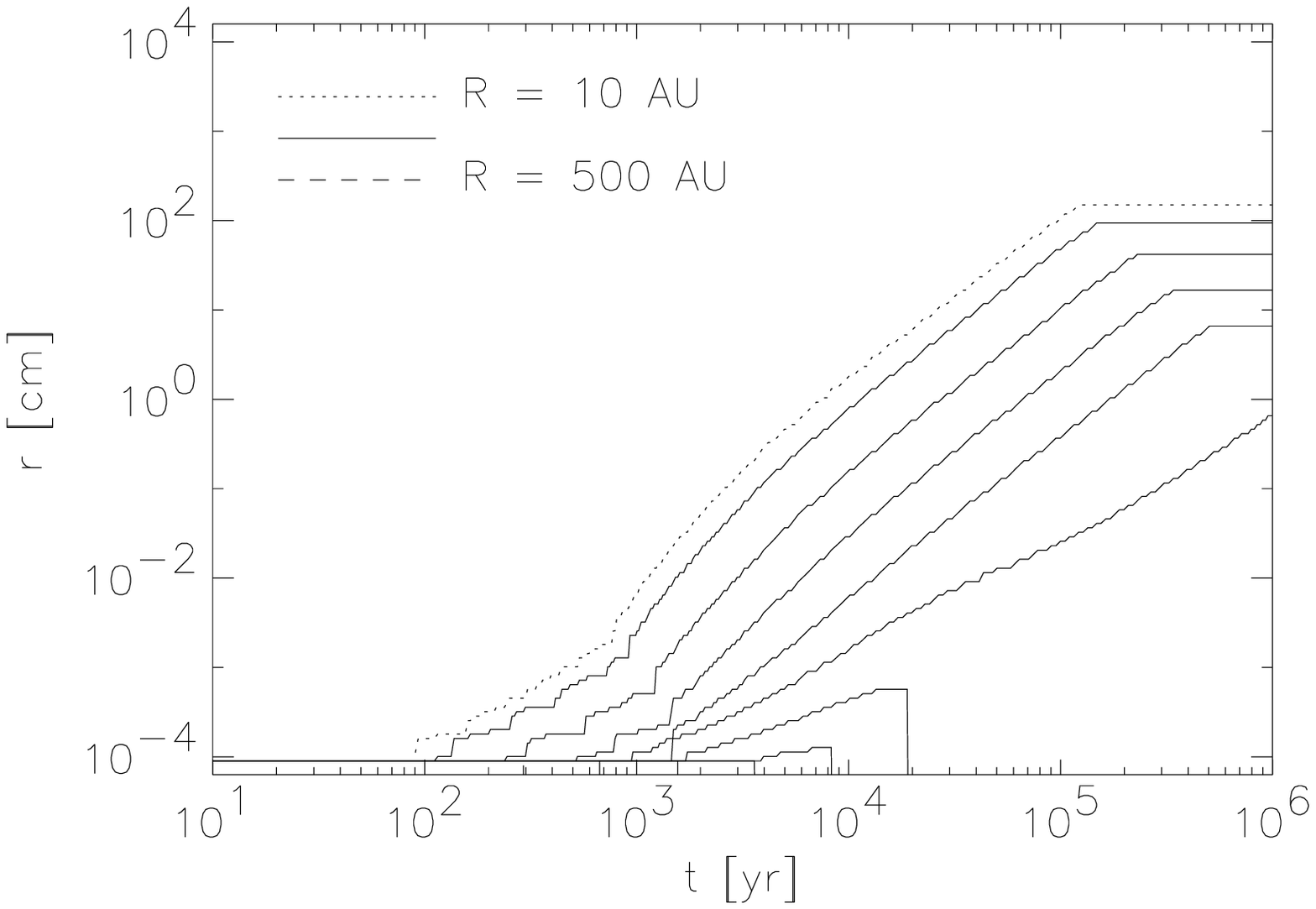}}\\}
\vskip -0.2 in

\centerline{\scalebox{0.65}{\includegraphics{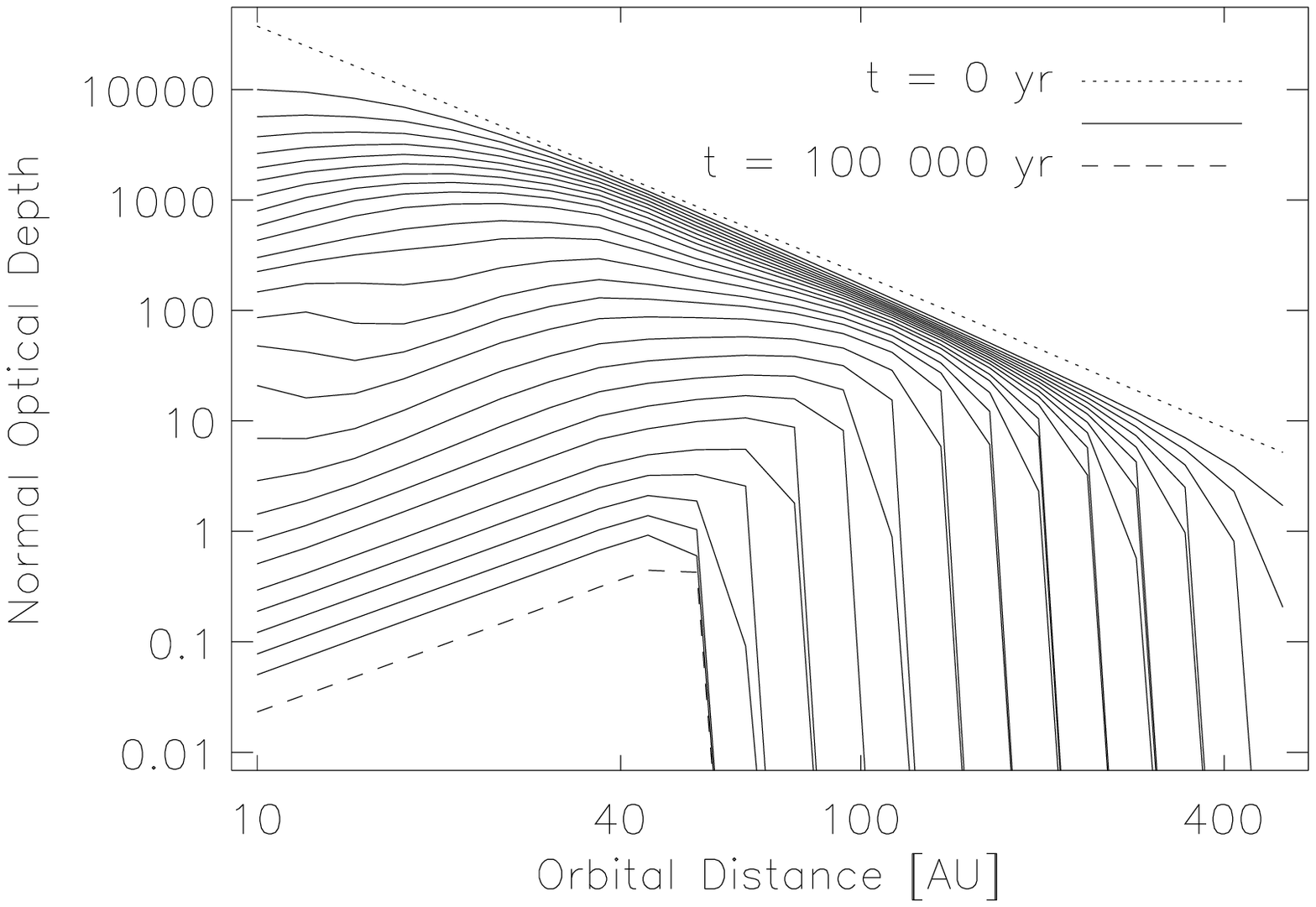}}}

\end{document}